\documentclass[12pt]{article}
\usepackage{graphics,epsf,epsfig}
\usepackage{amssymb}
\usepackage{bm}
\usepackage{natbib}

\begin{document}

\title{Stability of parallel/perpendicular domain boundaries in lamellar block
  copolymers under oscillatory shear}
\author{Zhi-Feng Huang and Jorge Vi\~nals \\
McGill Institute for Advanced Materials and Department of Physics,\\
McGill University, Montreal, QC H3A 2T8, Canada}

\maketitle

\begin{center}
\textbf{Synopsis}
\end{center}

We introduce a model constitutive law for the dissipative stress tensor of
lamellar phases to account for low frequency and long wavelength flows. Given
the uniaxial symmetry of these phases, we argue that the stress tensor must be
the same as that of a nematic but with the local order parameter being the slowly
varying lamellar wavevector. This assumption leads to a dependence of the
effective dynamic viscosity on orientation of the lamellar phase. We then consider
a model configuration comprising a domain boundary separating laterally unbounded
domains of so called parallel and perpendicularly oriented lamellae in a
uniform, oscillatory, shear flow, and show that the configuration can be
hydrodynamically unstable for the constitutive law chosen. It is argued that
this instability and the secondary flows it creates can be used to infer
a possible mechanism for orientation selection in shear experiments.

\newpage

\section{INTRODUCTION}

Recent interest in block copolymers arises from their ability to
self-assemble at the nanoscale through microphase
separation and ordering, leading to 
mesophases with various types of symmetries, such as lamellar,
cylindrical, or spherical [\citet{re:fredrickson96,re:larson99}]. 
However, when processed by thermal quench or
solvent casting from an isotropic, disordered state, a macroscopic sample
manifests itself as a polycrystalline configuration consisting of
locally ordered but randomly oriented domains (or grains), with the
presence of large amount of topological defects and unusual 
rheological properties. The development of the equilibrium
state characterized by macroscopic orientational order, as desired
in most of applications, requires unrealistically long times; hence 
external forces, such as steady or oscillatory shears
are usually applied to accelerate domain coarsening
and induce long range order. However, the mechanisms responsible
for the response of the copolymer microstructure and the selection of
a particular orientation over a macroscopic scale are still poorly
understood. In this paper we focus on the case of imposed oscillatory shear
flows on lamellar phases of block copolymers, and present an orientation
selection mechanism originating from an effective viscosity contrast
between lamellar phases of different orientation. Our study is based on a
mesoscopic or coarse-grained description of a copolymer, but our
results are expected to apply to other systems with the same symmetry
as a microphase separated block copolymer.

\begin{figure}
\centerline{\epsfig{figure=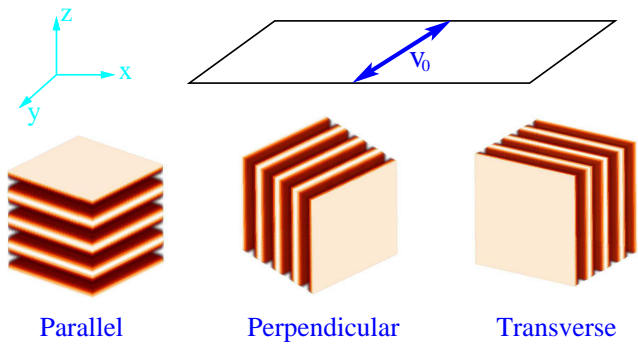,width=3.5in}}
\caption{Three lamellar orientations (Parallel, Perpendicular, and
  Transverse) under shear flow.}
\label{fig_ori}
\end{figure}

The response of lamellar block copolymers to
external shear flow can be classified according to three possible
uniaxial orientations (see 
Fig. \ref{fig_ori}): parallel (with lamellar planes parallel to the
shearing surface), perpendicular (with lamellae normal along the
vorticity direction of the shear flow), and transverse (with lamellae
normal directed along the shear). 
Solid-like or elastic response is expected for lamellar phases of
transverse orientation, while fluid-like or viscous response follows for the
other two, leading to different rheological properties at low
shear frequencies as measured experimentally 
[\citet{re:koppi92,re:fredrickson96}]. It is known that both
parallel and perpendicular alignments are favored over the transverse
one under shear, and either of them would be ultimately selected by
shear flow, as observed in most shear aligning experiments 
[\citet{re:larson99}]
(although some of experimental work indicates a coexistence between
parallel and transverse orientations [\citet{re:pinheiro96}],
fact that might be a result of strong segregation and/or molecular
entanglement). Of particular interest, and the least understood, is
the selection between parallel and
perpendicular orientations and the dependence on shear 
frequency $\omega$ and strain amplitude $\gamma$ 
[\citet{re:koppi92,re:maring97}] as well as
temperature [\citet{re:koppi92,re:pinheiro98}]. Near the
order-disorder transition temperature $T_{\rm ODT}$ and at low
shear frequencies, parallel alignment has been found in
poly(ethylene-propylene)-poly-(ethylethylene) (PEP-PEE) samples
[\citet{re:koppi92}], while for poly(styrene)-poly-(isoprene) (PS-PI)
copolymers, the observed ultimate orientation is parallel 
[\citet{re:maring97,re:leist99}]
or perpendicular [\citet{re:patel95}], depending on sample processing
details such as thermal history and shear starting time [\citet{re:larson99}].
At higher but still intermediate frequencies (where $\omega<\omega_c$, with
$\omega_c$ the characteristic frequency of polymer chain relaxation
dynamics), the preferred orientation is perpendicular for both PEP-PEE
and PS-PI copolymers under high enough shear strain 
[\citet{re:koppi92,re:patel95,re:maring97,re:leist99}]. At high
frequencies ($\omega>\omega_c$) the
orientation selected is different for PEP-PEE (perpendicular) than
PS-PI (parallel).

No basic understanding exists about the mechanisms underlying the above
complex phenomenology of orientation selection despite intense
theoretical scrutiny in recent years. For the frequency range
$\omega<\omega_c$ that we are interested in here, the detailed
relaxation dynamics of the polymer chains within each block are not expected
to be important; thus one adopts a coarse-grained, reduced
description in terms of the local monomer density as the order parameter 
[\citet{re:leibler80,re:ohta86,re:fredrickson87}]. Most of the early
analyses of shear alignment of copolymers relied on thermal fluctuation
effects near the \textit{transition point} $T_{\rm ODT}$, and focused on the role of
\textit{steady shears}. A study by \citet{re:cates89} indicates that in the
vicinity of the order-disorder transition, steady shear would suppress
critical fluctuations in an anisotropic manner, increase the transition
temperature, and favor the perpendicular orientation. Further extension by
\citet{re:fredrickson94} by incorporating viscosity contrast
between the microphases has shown that the perpendicular alignment
becomes prevalent for high shear rates, while the parallel one would be
favored at low shear rates. 

Later consideration was given to situations that are not fluctuation dominated 
(such as well-aligned lamellar phases or defect structures), 
regarding both stability and defect dynamics. Stability differences between
uniform parallel and perpendicular structures subjected to \textit{steady
shears} have been found [\citet{re:goulian95}]
through the consideration of anisotropic viscosities in a uniaxial
fluid. Recent molecular dynamics studies as well as hydrodynamic analyses of
the smectic A phase [\citet{re:soddemann04,re:guo06}] 
have shown an undulation instability of parallel lamellae, as well as a
transition from fully ordered parallel to perpendicular phases for large
enough shear rate. Regarding the effect of \textit{oscillatory shears},
an analysis of secondary instabilities [\citet{re:drolet99,re:chen02}]
has shown that the extent of the stability region for the perpendicular
orientation is always larger than that of the parallel direction, and
as expected, both much larger than the transverse region. 
Importantly, the role of viscosity difference between the polymer
blocks (as introduced by \citet{re:fredrickson94} to address
shear effects near the transition point) is found to be
negligible for the stability of well-aligned lamellar structures, due
to its weak coupling to long wavelength perturbations. In order to
address experimental phenomena related to dynamics of
structure evolution and orientation selection, more recent
theoretical efforts focused on the dynamic competition between coexisting
phases of different orientations. Examples include the study
of a grain boundary separating parallel and transverse lamellar
domains under oscillatory shears that described the dependence of the grain
boundary velocity on shearing parameters such as frequency
and amplitude [\citet{re:huang03,re:huang04}].

However, we are still far from accounting for
existing experimental phenomenology on orientation selection, possibly
because current approaches and models for block
copolymer dynamics might not be adequate. It is important to note that
block copolymer viscous response is not Newtonian even in the limit of
vanishing frequency $\omega \rightarrow 0$, since Newtonian
response would result in the degeneracy of parallel and perpendicular
orientations, contrary to experimental findings. In the theory of
\citet{re:fredrickson94}, the Newtonian assumption is used for
individual microphases, with different Newtonian viscosities chosen for
different monomers (blocks). We adopt here an alternative approach appropriate
for flows on a scale much larger than the lamellar spacing and address the
resulting deviation from Newtonian response in the low frequency limit. 
We introduce a constitutive law for the viscous stress tensor that explicitly
incorporates the uniaxial character of lamellar phases in analogy with
similar treatments of anisotropic fluids [\citet{re:ericksen60,re:leslie66}]
and nematic liquid crystals
[\citet{re:forster71,re:martin72,re:degennes93}]. As will be
shown below, the resulting effective viscosity depends on
lamellar orientation, in qualitative agreement with experiment results.

The focus of our analysis is the competition among coexisting but differently
oriented lamellar domains under oscillatory shears 
in a polycrystalline sample. 
We are primarily concerned with a simplified configuration involving two
regions of parallel and perpendicular orientations, and use our
results to address the selection mechanism in multi domain configurations.
It will be argued below that domain competition, and the orientation
selection mechanism it provides, could be determined by viscosity
contrast between lamellar phases due to their different 
orientations with respect to the imposed shear. A
hydrodynamic instability is predicted at the interface separating
parallel and perpendicular lamellae for certain ranges of material and
shearing parameters. The resulting nonuniform
secondary flows are argued to lead to the perpendicular orientation in some
ranges of parameters. Comparison of our results to existing experimental findings,
as well as possible examination of our predictions, are
also discussed.

\section{MODEL}

\subsection{Governing equations}
\label{subsec_hydro}

In the range of low shear frequencies compared to the inverse of the
polymer chain relaxation time, the 
phase behavior of a block copolymer is determined by an order parameter
field $\psi$ representing the variation of local monomer density, and
a velocity field ${\mathbf v}$. Evolution of $\psi$ is governed by a
time-dependent Ginzburg-Landau equation
\begin{equation}
\partial \psi / \partial t + {\mathbf v} \cdot {\bm \nabla} \psi
= - \Lambda \delta {\cal F} / \delta \psi,
\label{eq_G-L}
\end{equation}
with ${\cal F}$ the coarse-grained free energy given by
\citet{re:leibler80} and \citet{re:ohta86}, and $\Lambda$ an Onsager
kinetic coefficient. Equation (\ref{eq_G-L}) is coupled to the
following equation governing local velocity field 
${\mathbf v}=(v_x,v_y,v_z)$ for a block copolymer
system under oscillatory shear flow:
\begin{equation}
Re \left ( \partial {\mathbf v} / \partial t + {\mathbf v} \cdot
  {\mathbf \nabla} {\mathbf v} \right ) = - {\mathbf \nabla} p 
  + {\mathbf \nabla} \cdot {\bm \sigma}^D,
\label{eq_N-S}
\end{equation}
with the incompressibility condition ${\mathbf \nabla} \cdot {\mathbf v} = 0$.
Here $p$ is the pressure field, and $Re$ is the Reynolds number defined as
\begin{equation}
Re = \rho \omega d^2 / \eta
\label{eq_Re}
\end{equation}
with $\omega$ the shear frequency, $d$ the thickness of copolymer
system confined between shear planes, $\rho$ the copolymer
density, and $\eta$ a Newtonian viscosity. The coupling in
Eqs. (\ref{eq_G-L}) and (\ref{eq_N-S}) is complex,
especially when the stress tensor ${\bm \sigma}^D$ in Eq. (\ref{eq_N-S})
depends on the concentration field $\psi$, and when fluid
inertia cannot be ignored (i.e., at nonzero Reynolds number). In order to
simplify our analysis, we will assume that order parameter diffusion in
Eq. (\ref{eq_G-L}) is negligible, so that $\psi$ is advected by the flow
${\mathbf v}$. At the end of the analysis, we will discuss possible implications of
the flow fields obtained on order parameter diffusion as given by
Eq. (\ref{eq_G-L}). 

Equation (\ref{eq_N-S}) has been made dimensionless by introducing a length scale 
$d$, a time scale $\omega^{-1}$, and by rescaling pressure $p
\rightarrow p/(\eta \omega)$. In addition to a Newtonian viscous term,
Eq. (\ref{eq_N-S}) includes a dissipative stress tensor
${\bm \sigma}^D$ appropriate for a phase with
uniaxial symmetry. We assume the constitutive equation (in
dimensionless form)
due to Ericksen [\citet{re:ericksen60,re:leslie66}] as 
originally derived for anisotropic fluids and later applied to a nematic
liquid crystal
\begin{equation}
\sigma^D_{ij} = D_{ij} + \alpha_1 \hat{n}_i \hat{n}_j \hat{n}_k
\hat{n}_l D_{kl} + \alpha_{56} (\hat{n}_i \hat{n}_k D_{jk} + \hat{n}_j
\hat{n}_k D_{ik}),
\label{eq_E-L}
\end{equation}
where $D_{ij}=\partial_i v_j + \partial_j v_i$ ($i,j=x,y,z$), 
but with $\hat{\mathbf n}=(\hat{n}_x,\hat{n}_y,\hat{n}_z)$ a 
unit vector defining the \textit{slowly varying} local normal
of the lamellar phase. 
Eq. (\ref{eq_E-L}) is expected to be generic for any uniaxial phase,
and thus should apply to the lamellar phases studied here by reason
of symmetry. There are two independent viscosities $\alpha_1$ and
$\alpha_{56}$ (the Leslie's coefficients in the notation used by 
\citet{re:degennes93}), which have been rescaled as
$$
\alpha_1 \rightarrow \alpha_1 / \eta \quad {\rm and} \quad
\alpha_{56} \rightarrow \alpha_{56} / \eta.
$$
They obey the
relationship (due to the positivity of entropy production 
[\citet{re:forster71,re:martin72}])
\begin{equation}
\alpha_{56} \ge -1 \quad {\rm and} \quad \alpha_1 +2\alpha_{56} \ge -1.
\label{eq_alpha}
\end{equation}

For a uniform lamellar phase of parallel orientation ($\hat{\mathbf
  n}=(0,0,1)$ for the imposed shear flow along the $y$ direction considered,
  see Fig. \ref{fig_ori}), the Navier Stokes equation can be written 
explicitly as
\begin{eqnarray}
Re \left ( \partial_t v_x + v_j \partial_j v_x \right ) &=& 
-\partial_x p + \partial_j^2 v_x + \alpha_{56} (\partial_z^2 v_x 
+\partial_x \partial_z v_z), \nonumber\\
Re \left ( \partial_t v_y + v_j \partial_j v_y \right ) &=& 
-\partial_y p + \partial_j^2 v_y + \alpha_{56} (\partial_z^2 v_y
+\partial_y \partial_z v_z), \label{eq_Paral}\\
Re \left ( \partial_t v_z + v_j \partial_j v_z \right ) &=& 
-\partial_z p + (1+\alpha_{56})\partial_j^2 v_z
+2(\alpha_1 + \alpha_{56})\partial_z^2 v_z, \nonumber
\end{eqnarray}
after substitution of Eq. (\ref{eq_E-L}) into Eq. (\ref{eq_N-S}). For a domain
of perpendicular orientation ($\hat{\mathbf n}=(1,0,0)$) we have instead
\begin{eqnarray}
Re \left ( \partial_t v_x + v_j \partial_j v_x \right ) &=& 
-\partial_x p + (1+\alpha_{56})\partial_j^2 v_x
+2(\alpha_1 + \alpha_{56})\partial_x^2 v_x, \nonumber\\
Re \left ( \partial_t v_y + v_j \partial_j v_y \right ) &=& 
-\partial_y p + \partial_j^2 v_y + \alpha_{56} (\partial_x^2 v_y
+\partial_x \partial_y v_x), \label{eq_Perpen}\\
Re \left ( \partial_t v_z + v_j \partial_j v_z \right ) &=& 
-\partial_z p + \partial_j^2 v_z + \alpha_{56} (\partial_x^2 v_z
+\partial_x \partial_z v_x). \nonumber
\end{eqnarray}
The Newtonian limit is recovered by setting 
$\alpha_1=\alpha_{56}=0$. Contrary to previous work on block
copolymers in the creeping flow approximation
[\citet{re:goulian95,re:fredrickson94,re:chen02}], we retain inertial
terms in the above equations. Although small, they are
significant in determining the instabilities discussed below.
In the regime of small $Re$, both uniform parallel and perpendicular configurations
are hydrodynamically stable as derived from Eqs. (\ref{eq_Paral}) and
(\ref{eq_Perpen}) with a procedure similar to that shown in
Sec. \ref{sec_analy} for the grain boundary configuration.

Our assumption, Eq. (\ref{eq_E-L}), is consistent with experimental
determinations of the loss modulus $G''$ and dynamic viscosity 
$\eta'$ ($=G''/\omega$) for uniform phases of different lamellar
orientations. Assuming a shear flow along $y$
(Fig. \ref{fig_ori}), we have in the creeping flow limit $Re \rightarrow 0$
(in dimensional form) 
\begin{itemize}
\item[-] {\it Perpendicular} ($\hat{\mathbf n}=(1,0,0)$): $\eta'=\eta$;
\item[-] {\it Parallel} ($\hat{\mathbf n}=(0,0,1)$):
  $\eta'=\eta+\alpha_{56}$;
\item[-] {\it Transverse} ($\hat{\mathbf n}=(0,1,0)$): 
  $\eta'=\eta+\alpha_{56}+\frac{2a^2}{(1+a^2)^2}\alpha_1$ ($a=\gamma
  \sin(\omega t)$, with $\gamma$ the shear strain amplitude); when averaged over
  period $T$, it is $\langle \eta' \rangle _T 
  =\eta+\alpha_{56}+\frac{\gamma^2}{(1+\gamma^2)^{3/2}} \alpha_1$.
\end{itemize}
When $\alpha_1>0$ and $\alpha_{56}>0$, we obtain $\eta'_{\rm Transverse}
> \eta'_{\rm Parallel} > \eta'_{\rm Perpendicular}$, in agreement with
low $\omega$ experimental results of \citet{re:koppi92} for PEP-PEE
diblock copolymers, and also with the result of molecular dynamics
simulations by \citet{re:guo06} showing smaller viscosity of
perpendicular lamellae compared to that of the parallel phase. 
On the other hand, when $\alpha_{56}<0$, we have
$\eta'_{\rm Parallel} < \eta'_{\rm Perpendicular}$, which is the case
in PS-PI copolymers [\citet{re:chen98b,re:gupta95}]. 
The viscosity coefficients $\alpha_1$ and $\alpha_{56}$ could be in
principle measured in block copolymers under shear flows, with
experimental setups possibly similar to those used in nematic liquid
crystals [see, e.g., methods reviewed by \citet{re:degennes93}].

\subsection{Parallel/perpendicular configuration and viscosity contrast}

\begin{figure}
\centerline{\epsfig{figure=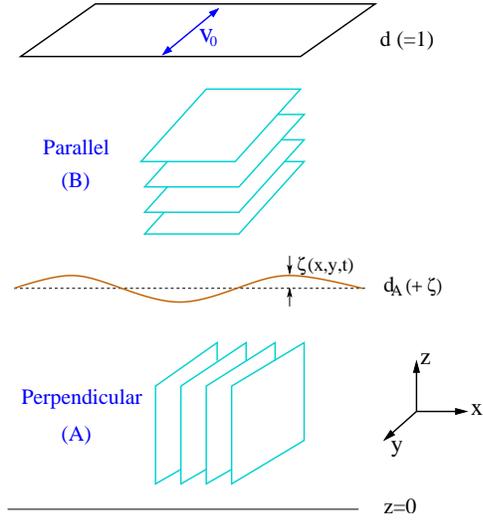,width=2.5in}}
\caption{A parallel/perpendicular configuration subjected to
  oscillatory shear flow.}
\label{fig_conf}
\end{figure}

We focus on coexisting lamellar domains of parallel and perpendicular orientations
under oscillatory shear flows. 
In real samples these two types of domains may be separated by topological 
defects such as grain boundaries, dislocations, or disclinations. We
study here the simplified configuration shown in Fig. \ref{fig_conf}, 
comprising two fully ordered, three-dimensional lamellar domains that
are identical except for their orientation and thickness. 
The perpendicular domain A
is of thickness $d_A$, and the parallel domain B of $d_B=1-d_A$ (in
dimensionless form), both confined between a pair of shearing planes.
The system is uniformly sheared along the $y$ direction; the velocity
field ${\mathbf v}$ is zero on the lower boundary plane $z=0$
and equal to ${\mathbf v}_0=\gamma \cos t ~\hat{\mathbf y}$ (or $\gamma \omega d
\cos(\omega t) ~\hat{\mathbf y}$ in dimensional form) on the upper 
boundary $z=1$. 

Under the shear considered here, bulk configurations of either orientation
(parallel or perpendicular) are linearly stable. However, we will show that
the configuration of Fig. \ref{fig_conf} can become unstable.
Note that the effective viscosities of the two domains are different,
as discussed in Sec. \ref{subsec_hydro} Therefore this configuration
is analogous to the case of two superposed Newtonian fluids of
different viscosity, which is known to be unstable under steady (plane
Couette flow [\citet{re:yih67,re:hooper85}]) or oscillatory
[\citet{re:king99}] shears. In the present case, however, the viscosity
contrast follows from the orientation dependence of our constitutive
law (\ref{eq_E-L}). This contrast between the lamellar
phases can cause interface instability, but with a dependence
on shearing conditions and system parameters much more complicated
than that of the Newtonian limit.

\section{HYDRODYNAMIC STABILITY ANALYSIS}
\label{sec_analy}

\subsection{Base flow}

The base state for the configuration of Fig. \ref{fig_conf} is 
a planar interface located at $z=d_A$,
separating two stable perpendicular (A) and parallel (B) regions.
Under uniform shear the velocity fields are along
the $y$ direction: ${\mathbf v}_{A,B}=(0,V_{A,B},0)$. 
In dimensionless form, the velocities are given by 
\begin{eqnarray}
V_A &=& {\cal R}e \left \{ \alpha_A \ \sinh[(1+i)\beta_A z] \ \gamma e^{it}
\right \}, \nonumber\\
V_B &=& {\cal R}e \left \{ \alpha_A \ \left ( \sinh[(1+i)\beta_A d_A] \
\cosh[(1+i)\beta_B (z-d_A)] \right. \right.  \nonumber\\
 && + \sqrt{m} \left. \left. \cosh[(1+i)\beta_A d_A] \
\sinh[(1+i)\beta_B (z-d_A)] \ \right ) \gamma e^{it} \right \},
\label{eq_VAB}
\end{eqnarray}
with 
$$
\alpha_A^{-1} = \sinh[(1+i)\beta_A d_A] \ \cosh[(1+i)\beta_B d_B]
  + \sqrt{m} \cosh[(1+i)\beta_A d_A] \ \sinh[(1+i)\beta_B d_B]
$$
and the viscosity ratio $m=\mu_A/\mu_B$ (with $\mu_A=1$ for the
perpendicular region and $\mu_B=1+\alpha_{56}$ for a parallel
domain). Here $\beta_A$ and $\beta_B$ are the inverse Stokes layer
thicknesses
\begin{equation}
\beta_{A,B} = \left ( \frac{Re}{2\mu_{A,B}} \right )^{1/2}.
\label{eq_beta}
\end{equation}
Also for the pressure field, $p_A=p_B=p_0$. Note that these base
state solutions are the same as those of two superposed Newtonian
fluids with different viscosities $\mu_A$ and $\mu_B$ under oscillatory
Couette flow [\citet{re:king99}].

\subsection{Perturbation analysis}

For Newtonian fluids or in some viscoelastic models (e.g.,
Oldroyd-B or Maxwell fluids), the three-dimensional stability problem 
can be reduced to an effective two-dimensional one, with fluid
stability under shear flows governed by Orr-Sommerfeld type equations
for a single stream function 
describing two-dimensional disturbances. However, this is not the
case discussed here and governed by Eqs. (\ref{eq_N-S}) and (\ref{eq_E-L}), as will
be seen below.

We expand both velocity and pressure fields into the base state
given above and perturbations,
\begin{equation}
v_i^{A,B} = V_{A,B} \delta_{iy} + u_i^{A,B} \; (i=x,y,z),
\qquad p_{A,B} = p_0 + p'_{A,B}.
\label{eq_u_p}
\end{equation}
These flow perturbations are accompanied by an
undulation of interface, denoted as $\zeta(x,y,t)$ (see
Fig. \ref{fig_conf}). The boundary conditions at $z=d_A +
\zeta(x,y,t)$ include:

Continuity of velocity
\begin{equation}
{\mathbf v}^{A} = {\mathbf v}^{B},
\label{eq_i1}
\end{equation}
continuity of tangential stress
\begin{eqnarray}
&&\left \{ [1-(\partial_x \zeta)^2] \sigma_{xz}^D - \partial_x \zeta 
\ \partial_y \zeta \ \sigma_{yz}^D - \partial_y \zeta \ \sigma_{xy}^D
+ (\sigma_{zz}^D - \sigma_{xx}^D) \partial_x \zeta \right \}_A^B =0,
\label{eq_i2} \\
&&\left \{ [1-(\partial_y \zeta)^2] \sigma_{yz}^D - \partial_x \zeta 
\ \partial_y \zeta \ \sigma_{xz}^D - \partial_x \zeta \ \sigma_{xy}^D
+ (\sigma_{zz}^D - \sigma_{yy}^D) \partial_y \zeta \right \}_A^B =0,
\label{eq_i3}
\end{eqnarray}
and balance of normal stress
\begin{equation}
\left \{ -p + \sigma_{zz}^D - \partial_x \zeta \ \sigma_{xz}^D 
- \partial_y \zeta \ \sigma_{yz}^D \right \}_A^B = -\Gamma' 
(\partial_x^2 \zeta + \partial_y^2 \zeta),
\label{eq_i4}
\end{equation}
where $\{ \}_A^B = \{ \}_B - \{ \}_A$, $\sigma_{ij}^D$ is the
dissipative stress tensor, and 
\begin{equation}
\Gamma' = \Gamma / (\eta \omega d),
\label{eq_G}
\end{equation}
with $\Gamma$ the interfacial tension.
Also, the kinematic condition at the interface yields
\begin{equation}
\left ( \partial_t + {\mathbf v}^B \cdot {\mathbf \nabla} \right )
\zeta = v_z^B.
\label{eq_i5}
\end{equation}

Assume expansions of the form
\begin{equation}
u_i^{A,B} = \sum \limits_{q_x,q_y} \hat{u}_i^{A,B}(q_x,q_y,z,t)
\exp \left [ i(q_x x + q_y y) \right ],
\label{eq_u}
\end{equation}
where $q_x$ and $q_y$ are wave numbers in the $x$ and $y$ directions.
Substituting (\ref{eq_u}) into Eqs. (\ref{eq_Paral}) and
(\ref{eq_Perpen}), retaining terms up to first order in the perturbation
amplitudes, and eliminating 
the pressure, we obtain the following equations for perturbed
velocity fields $\hat{u}_z^{A,B}$ and $\hat{u}_x^{A,B}$, which govern the
system stability when combined with the above interfacial conditions:

For the parallel region B ($d_A \le z \le 1$),
\begin{eqnarray}
& Re \left [ (\partial_t + iq_y V_B)(\partial_z^2 - q^2) \hat{u}_z^{B}
 - iq_y(\partial_z^2 V_B) \hat{u}_z^{B} \right ] & \nonumber\\
& =(1+\alpha_{56})
(\partial_z^2 - q^2)^2 \hat{u}_z^{B} - 2\alpha_1 q^2 \partial_z^2 
\hat{u}_z^{B}, & \label{eq_uz_B}
\end{eqnarray}
\begin{eqnarray}
&& Re \left [ \partial_t(\partial_z^2 - q^2) \hat{u}_x^{B} +
  iq_y(\partial_z^2 - q^2)(V_B \hat{u}_x^{B}) + 
  2q_xq_y(\partial_z V_B) \hat{u}_z^{B} \right ] \nonumber\\
&& = (\partial_z^2 - q^2)^2 \hat{u}_x^{B} +\alpha_{56}
(\partial_z^2 - q^2) \left ( \partial_z^2 \hat{u}_x^{B}
-iq_x \partial_z \hat{u}_z^{B} \right ) -2iq_x \alpha_1 
\partial_z^3 \hat{u}_z^{B}, \label{eq_ux_B}
\end{eqnarray}
(here $q^2=q_x^2 + q_y^2$), while for the perpendicular region A 
($0 \le z \le d_A$),
\begin{eqnarray}
&& Re \left [ (\partial_t + iq_y V_A)(\partial_z^2 - q^2) \hat{u}_z^{A}
 - iq_y(\partial_z^2 V_A) \hat{u}_z^{A} \right ]  \nonumber\\
&& = (\partial_z^2 - q^2)^2 \hat{u}_z^A - \alpha_{56} q_x^2
(\partial_z^2 - q^2) \hat{u}_z^A + iq_x [2\alpha_1 q_x^2 - \alpha_{56}
(\partial_z^2 - q^2)] \partial_z \hat{u}_x^A, 
\label{eq_uz_A}
\end{eqnarray}
\begin{eqnarray}
&& Re \left [ \partial_t(\partial_z^2 - q^2) \hat{u}_x^{A} +
  iq_y(\partial_z^2 - q^2)(V_A \hat{u}_x^{A}) + 
  2q_xq_y(\partial_z V_A) \hat{u}_z^{A} \right ] \nonumber\\
&& = (1+\alpha_{56})(\partial_z^2 - q^2)^2 \hat{u}_x^{A}
  - 2\alpha_1 q_x^2 (\partial_z^2 - q_y^2) \hat{u}_x^{A},
\label{eq_ux_A}
\end{eqnarray}
with rigid boundary conditions on the planes $z=0$ and $z=1$
\begin{eqnarray}
& \hat{u}_x^A(0) = \hat{u}_z^A(0) = \partial_z \hat{u}_x^A(0) 
= \partial_z \hat{u}_z^A(0) = 0, & \nonumber\\
& \hat{u}_x^B(1) = \hat{u}_z^B(1) = \partial_z \hat{u}_x^B(1) 
= \partial_z \hat{u}_z^B(1) = 0. & \label{eq_bound}
\end{eqnarray}
The other velocity field $\hat{u}_y^{A,B}$ can be
obtained from the incompressibility condition.
Note that the form of Eqs. (\ref{eq_uz_B})--(\ref{eq_ux_A}) is similar
to that of the Orr-Sommerfeld equation for Newtonian fluids; 
however, in the above equations velocity fields are coupled 
(except for $q_x=0$), and thus the three-dimensional system here
cannot be reduced to an effective two-dimensional one described by
only one stream function. This irreducibility might be understood from
the fact that parallel and perpendicular orientations can be
distinguished only in three-dimensional space, and thus the results of
the associated flows that are orientation dependent should be also
three-dimensional.

The solutions to the above problem can be found by writing
\begin{eqnarray}
& \hat{u}_z^{A,B}(z,t) = e^{\sigma t} \phi_z^{A,B}(z,t), & \nonumber\\
& \hat{u}_x^{A,B}(z,t) = e^{\sigma t} \phi_x^{A,B}(z,t), &
\label{eq_floquet}\\
& \hat{\zeta}(t) = e^{\sigma t} h(t), & \nonumber
\end{eqnarray}
with interfacial perturbation $\hat{\zeta}$ defined by 
$\zeta(x,y,t)=\sum_{q_x,q_y} \hat{\zeta}(t) \exp[i(q_x x+q_y y)]$.
When $q_y \neq 0$, according to Floquet's theorem $\phi_{z,x}^{A,B}$
and $h$ are periodic in time $t$ with period $T$ ($=2\pi$ here) 
if the eigenvalue $\sigma$ (the Floquet exponent) is simple
[\citet{re:yih68}], since coefficients in
Eqs. (\ref{eq_uz_B})--(\ref{eq_ux_A}) that are proportional to the base flow
$V_{A,B}$ are periodic in $t$ as shown in Eq. (\ref{eq_VAB}). 
When $q_y=0$, $\phi_{z,x}^{A,B}$
and $h$ are time independent, and $\sigma$ represents the perturbation
growth rate. In either case, the real part of $\sigma$ determines the
system stability.

\subsection{Small Reynolds number limit}
\label{subsec_sol}

The Newtonian viscosity $\eta$ is very large for block copolymers,
resulting in small Reynolds numbers (Eq. (\ref{eq_Re})). For a
typical block copolymer of density $\rho \sim 1$ g cm$^{-3}$ 
and $\eta \sim 10^4-10^6$ P, $Re/\omega \sim 10^{-4}-10^{-6}$
s for a thickness $d \sim 1$ cm.
Within a reasonable range of frequencies $\omega$ of interest,
we have $Re \ll 1$. The stability problem can be now solved analytically 
by expanding around small Reynolds number, 
\begin{eqnarray}
& \phi_{z,x}^{A,B}(z,t) = \phi_{(z,x)0}^{A,B}(z,t) + 
Re \ \phi_{(z,x)1}^{A,B}(z,t) + ...,& \nonumber\\
&h = h_0(t) + Re \ h_1(t) + ...,& \label{eq_expan}\\
&\sigma = \sigma_0 + Re \ \sigma_1 + ...,& \nonumber
\end{eqnarray}
with $\phi$ and $h$ functions defined in Eq. (\ref{eq_floquet}).
Note that the order of the rescaled
interfacial tension $\Gamma'$, defined in Eq. (\ref{eq_G})
is $\omega$ dependent, and needs to be addressed separately. 
In the limit of $Re \rightarrow 0$ with $\Gamma'$ finite, we
set $\Gamma'=\Gamma_0={\cal O}(1)$, while for small but finite
values of Reynolds numbers, $\Gamma'$ can be expressed as a power of $Re$. 
For a typical value of $\Gamma \sim 1$ dyne/cm (as estimated from the
interfacial energy between two blocks [\citet{re:helfand72,re:gido94}])
and large $\eta$ ($\sim
10^4-10^6$ P) appropriate for block copolymers, we have $\Gamma'\omega \sim
10^{-4}-10^{-6}$ s$^{-1}$. The intermediate range of $\omega$ (around
1 s$^{-1}$) for typical experiments leads to 
$\Gamma'\equiv Re~\Gamma_1={\cal O}(Re)$, 
whereas for much larger frequencies $\Gamma'/Re \ll 1$. 
In the following, we present the solutions of
zeroth order (containing $\Gamma_0$) and first order (containing $\Gamma_1$)
in $Re$, which are accurate enough to determine the stability behavior
for lamellar block copolymers with typical $Re \ll 1$.

\subsubsection*{C.1 $Re \rightarrow 0$ (with $\Gamma'=\Gamma_0={\cal O}(1)$)}

In this parameter range, we only need the zeroth-order solution for
the velocity fields, given by (for $q_x \neq 0$)
\begin{eqnarray}
\phi_{z0}^{B} &=& B_1^{(0)} e^{b_1 z} + B_2^{(0)} e^{b_2 z} 
+ B_3^{(0)} e^{b_3 z} + B_4^{(0)} e^{b_4 z}, \label{eq_phi_z0B}\\
\phi_{x0}^B &=& C_1^{(0)} e^{r_1 z} + C_2^{(0)} e^{r_2 z} 
+ C_3^{(0)} e^{r_3 z} + C_4^{(0)} e^{r_4 z}
+ \frac{iq_x}{q^2} \partial_z \phi_{z0}^B, \label{eq_phi_x0B}\\
\phi_{z0}^A &=& A_1^{(0)} e^{a_1 z} + A_2^{(0)} e^{a_2 z} 
+ A_3^{(0)} e^{a_3 z} + A_4^{(0)} e^{a_4 z} \nonumber\\
  &+& D_1^{(0)} A_1' e^{s_1 z} + D_2^{(0)} A_2' e^{s_2 z} 
   + D_3^{(0)} A_3' e^{s_3 z} + D_4^{(0)} A_4' e^{s_4 z},
\label{eq_phi_z0A}\\
\phi_{x0}^A &=& D_1^{(0)} e^{s_1 z} + D_2^{(0)} e^{s_2 z} 
+ D_3^{(0)} e^{s_3 z} + D_4^{(0)} e^{s_4 z}, \label{eq_phi_x0A}
\end{eqnarray}
with 
\begin{eqnarray}
&& b_i^2 = \frac{q^2}{\mu_B} \left \{ \mu_B + \alpha_1 \pm
\left [ \alpha_1 (2\mu_B + \alpha_1) \right ]^{1/2} \right \},
\label{eq_b}\\
&& s_i^2 = \left \{ \mu_B q^2 + \alpha_1 q_x^2 \pm q_x^2
\left [ \alpha_1 (2\mu_B + \alpha_1) \right ]^{1/2} \right \}
/ \mu_B, \label{eq_s}\\
&& A_i' = iq_x \left ( \frac{2\alpha_1 - \alpha_{56}^2}
{s_i^2 - q^2 -\alpha_{56} q_x^2} - \frac{\mu_B}{q_x^2} \right ) 
s_i \label{eq_A'}
\end{eqnarray}
for $i=1,2,3,4$, and 
\begin{eqnarray}
&& r_1 = q, \; r_2 = -q, \; r_3 = q/\sqrt{\mu_B}, \; r_4 =
-q/\sqrt{\mu_B}, \label{eq_r}\\
&& a_1 = q, \; a_2 = -q, \; a_3 = \sqrt{q^2+\alpha_{56}q_x^2}, \;
a_4 = -a_3. \label{eq_a}
\end{eqnarray}
For $q_x=0$, we have $\phi_{x0}^A = \phi_{x0}^B =0$, $\phi_{z0}^B$
is also given by Eq. (\ref{eq_phi_z0B}), and
\begin{equation}
\phi_{z0}^A = (A_1^{(0)} + A_2^{(0)} z) e^{qz} + 
(-A_1^{(0)} + A_3^{(0)} z) e^{-qz}. \label{eq_phi_z0A_qx0}
\end{equation}

By using the boundary conditions (\ref{eq_bound}), we can express the
coefficients $A_{3,4}^{(0)}$, $B_{3,4}^{(0)}$, $C_{3,4}^{(0)}$, 
and $D_{3,4}^{(0)}$ in terms of $A_{1,2}^{(0)}$, $B_{1,2}^{(0)}$,
$C_{1,2}^{(0)}$, and $D_{1,2}^{(0)}$. The remaining 
coefficients are obtained by solving a linear matrix equation
\begin{equation}
{\bm \chi} \cdot {\bm A^{(0)}} = {\bm h^{(0)}} h_0,
\label{eq_chi0}
\end{equation}
where ${\bm A^{(0)}} = [A_j^{(0)},B_j^{(0)},C_j^{(0)},D_j^{(0)}]$
($j=1,2$) for $q_x \neq 0$ or $[A_j^{(0)},B_j^{(0)}]$ for $q_x=0$, 
${\bm \chi}$ is a $8 \times 8$ or $4 \times 4$ constant matrix 
determined by interfacial conditions
(\ref{eq_i1})--(\ref{eq_i4}), $h_0=h_0(t)$ denotes the $0th$-order
interface perturbation, and the matrix ${\bm h^{(0)}}$ is a function of
$\Gamma_0$ and $(\partial_z V_B)_0$, with
$(\partial_z V_B)_0$ the gradient of $0th$-order base flow $V_{B}^{(0)}$:
\begin{equation}
V_{B}^{(0)} = [\lambda_0 d_A + \delta (z-d_A)] \gamma \cos t,
\quad (\partial_z V_B)_0 = \delta \gamma \cos t,
\label{eq_dz_VB0}
\end{equation}
with
$$
\delta = m \lambda_0,
\quad \lambda_0 = (d_A + m d_B)^{-1} \quad (m=\mu_A/\mu_B).
$$

To lowest order, the kinematic equation (\ref{eq_i5}) yields
$$
\partial_t h_0 = -(\sigma_0 + iq_y V_{B0}) h_0 + \phi_{z0}^B(d_A),
$$
where $V_{B0}=V_{B}^{(0)}(z=d_A)=\lambda_0 d_A \gamma \cos t$. 
From Eqs. (\ref{eq_phi_z0B}) and (\ref{eq_chi0}), $\phi_{z0}^B(d_A)$
can be expressed as 
\begin{equation}
\phi_{z0}^B(d_A) = \left [ f_{z0,1}^B(q_x,q_y) \Gamma_0 +
  f_{z0,2}^B(q_x,q_y) (\partial_z V_B)_0 \right ] h_0,
\label{eq_phiz0B_dA}
\end{equation}
with $f_{z0,1}^B$ and $f_{z0,2}^B$ complicated but known functions of
wave numbers $q_x$ and $q_y$ (and also dependent on parameters
$\alpha_1$, $\alpha_{56}$, and $d_A$), and hence
\begin{equation}
\partial_t h_0 = \left [ -\sigma_0 + f_{z0,1}^B(q_x,q_y) \Gamma_0
  \right ] h_0 + \left [ -iq_y V_{B0} + f_{z0,2}^B(q_x,q_y)
  (\partial_z V_B)_0 \right ] h_0.
\label{eq_h0_t}
\end{equation}
The requirement of periodicity of $h_0$ with time $t$ gives the 
$0th$-order Floquet exponent (note that terms proportional to
$V_{B0}$ and $(\partial_z V_B)_0$ are periodic in $t$)
\begin{equation}
\sigma_0 = f_{z0,1}^B(q_x,q_y) \Gamma_0.
\label{eq_sig0}
\end{equation}
Consequently, with initial value $h_0(0)$,
\begin{eqnarray}
h_0(t) &=& h_0(0) \exp \{ \int dt [ -iq_y V_{B0} + f_{z0,2}^B
  (\partial_z V_B)_0 ] \} \nonumber\\
  &=& h_0(0) \exp \left [ ( -iq_y d_A + m f_{z0,2}^B )
  \lambda_0 \gamma \sin t \right ], \label{eq_h0}
\end{eqnarray}
and the velocity fields are obtained from Eqs. (\ref{eq_floquet}), 
(\ref{eq_phi_z0B})--(\ref{eq_phi_x0A}), and (\ref{eq_chi0}), 
all proportional to $h_0(t)$.

\subsubsection*{C.2 $Re \ll 1$, $\Gamma' \ll 1$, such that $\Gamma'/Re
  = \Gamma_1 = {\cal O}(1)$}

In this case, the results of first order expansion need to be addressed.
When $q_x \neq 0$, the solution to ${\cal O}(Re)$ yields
\begin{eqnarray}
\phi_{z1}^B &=& \sum\limits_{i=1}^{4} \left [ B_i^{(1)} e^{b_i z} 
+B'_i z e^{b_i z} + B'_{i+4} z^2 e^{b_i z} \right ], \label{eq_phi_z1B}\\
\phi_{x1}^B &=& \sum\limits_{i=1}^{4} C_i^{(1)} e^{r_i z} 
+ \frac{iq_x}{q^2} \partial_z \phi_{z1}^B \nonumber\\ 
&+&\sum\limits_{i=1}^{4} C'_i z e^{r_i z}
+ \sum\limits_{i=3}^{4} C'_{i+2} z^2 e^{r_i z} 
+ \sum\limits_{i=1}^{4} B''_i e^{b_i z}, \label{eq_phi_x1B}\\
\phi_{z1}^A &=& \sum\limits_{i=1}^{4} \left [ A_i^{(1)} e^{a_i z}
+D_i^{(1)} A'_i e^{s_i z} + A'''_i z e^{a_i z} \right ] 
+ \sum\limits_{i=3}^{4} A'''_{i+2} z^2 e^{a_i z} \nonumber\\ 
&+&\sum\limits_{i=1}^{4} \left [ D''_i e^{s_i z} + D''_{i+4} z e^{s_i z}
  + D''_{i+8} z^2 e^{s_i z} \right ] \label{eq_phi_z1A}\\
\phi_{x1}^A &=& \sum\limits_{i=1}^{4} \left [ D_i^{(1)} e^{s_i z} 
+D'_i z e^{s_i z} + D'_{i+4} z^2 e^{s_i z} + A''_i e^{a_i z} 
\right ], \label{eq_phi_x1A}
\end{eqnarray}
while for $q_x=0$, we get $\phi_{x1}^A=\phi_{x1}^B=0$,
$\phi_{z1}^B$ given by Eq. (\ref{eq_phi_z1B}), and 
\begin{eqnarray}
\phi_{z1}^A &=& (A_1^{(1)} + A_2^{(1)} z) e^{qz} + 
(-A_1^{(1)} + A_3^{(1)} z) e^{-qz} \nonumber\\
 &+& (A'''_1 z^2 + A'''_2 z^3) e^{qz}
+ (A'''_3 z^2 + A'''_4 z^3) e^{-qz}. \label{eq_phi_z1A_qx0}
\end{eqnarray}

Here exponents $b_i$, $r_i$, $a_i$, $s_i$ and coefficients $A'_i$ are
the same as those in Eqs. (\ref{eq_b})--(\ref{eq_a}), and coefficients $B'_i$,
$B''_i$, $C'_i$, $A'_i$, $A''_i$, $A'''_i$, $D'_i$, and $D''_i$ are
complicated functions of $0th$ order solutions
$(A^{(0)},B^{(0)},C^{(0)},D^{(0)})$. The unknown first-order
coefficients $(A^{(1)},B^{(1)},C^{(1)},D^{(1)})$ can be determined by 
boundary conditions (\ref{eq_bound}) and interfacial conditions
(\ref{eq_i1})--(\ref{eq_i4}), as in the above $0th$ order case.
Similar to Eq. (\ref{eq_chi0}), the linear matrix equation governing
1st-order coefficients ${\bm A^{(1)}}$ ($=[A_j^{(1)},B_j^{(1)},
C_j^{(1)},D_j^{(1)}]$ for $q_x \neq 0$, or $[A_j^{(1)},B_j^{(1)}]$ 
for $q_x=0$, with $j=1,2$) is
\begin{equation}
{\bm \chi} \cdot {\bm A^{(1)}} = {\bm h^{(1)}} h_0 + {\bm h^{(0)}} h_1,
\label{eq_chi1}
\end{equation}
with matrices ${\bm \chi}$ and ${\bm h^{(0)}}$ the same as those in
Eq. (\ref{eq_chi0}), $h_1(t)$ the first-order interface perturbation,
and the matrix ${\bm h^{(1)}}$ a complicated function of $\Gamma_1$, 
$(\partial_z V_B)_0$, and $\partial_t (\partial_z V_B)_0$. 
Thus, the solution of Eq. (\ref{eq_chi1}) takes the form
\begin{equation}
{\bm A^{(1)}} = \left [ {\bm \chi}^{-1} {\bm h^{(1)}} \right ] h_0
+ \left [ {\bm A^{(0)}} / h_0 \right ] h_1, \label{eq_A1}
\end{equation}
with ${\bm A^{(0)}}$ given by
Eq. (\ref{eq_chi0}), and accordingly the first-order velocity fields 
can be calculated with the use of Eqs. 
(\ref{eq_phi_z1B})-(\ref{eq_phi_z1A_qx0}).

The ${\cal O}(Re)$ result of the kinematic condition (\ref{eq_i5}) is
given by
\begin{equation}
\partial_t h_1 = -(\sigma_0 + iq_y V_{B0}) h_1 
 -(\sigma_1 + iq_y V_{B1}) h_0 + \phi_{z1}^B(d_A),
\label{eq_h1_t}
\end{equation}
where $V_{B1}=\lambda_2 \gamma \sin t$ is the first order base flow
$V_{B}^{(1)}$ evaluated at interface $z=d_A$, with 
$$
\lambda_2 = \frac{1}{2} (\lambda_1 d_A-\lambda_0 d_A^3/3), \quad
\lambda_1 = [ (d_A^3 + m^2 d_B^3)/3 + m d_A d_B d] \lambda_0^2.
$$
The value of $\phi_{z1}^B(d_A)$ is determined by solution
(\ref{eq_phi_z1B}), and from Eq. (\ref{eq_A1}) we find
\begin{eqnarray}
&\phi_{z1}^B(d_A) = \left [ f_{z1,1}^B(q_x,q_y) \Gamma_1
 + f_{z1,2}^B(q_x,q_y)(\partial_z V_B)_0 
 +f_{z1,3}^B(q_x,q_y)(\partial_z V_B)_0^2
\right. & \nonumber\\
& \left. + f_{z1,4}^B(q_x,q_y)
\partial_t (\partial_z V_B)_0 \right ] h_0 + \left [ \phi_{z0}^B(d_A) / h_0 
\right ] h_1,&
\label{eq_phiz1B_dA}
\end{eqnarray}
with $\phi_{z0}^B(d_A)$ the $0th$-order solution as in
Eq. (\ref{eq_phiz0B_dA}), $f_{z1,1}^B=f_{z0,1}^B$, and $f_{z1,i}^B$
($i=2,3,4$) obtained from first-order solution (\ref{eq_A1}) and
functions of $\alpha_1$, $\alpha_{56}$, and $d_A$.
Substituting Eqs. (\ref{eq_phiz1B_dA}), (\ref{eq_dz_VB0}), and 
(\ref{eq_sig0}) into (\ref{eq_h1_t}), and using the condition that
$h_1$ is periodic in time, we obtain the first order Floquet
exponent
\begin{equation}
\sigma_1 = f_{z1,1}^B(q_x,q_y) \Gamma_1 
+ \frac{1}{2} \delta^2 \gamma^2 f_{z1,3}^B(q_x,q_y),
\label{eq_sig1}
\end{equation}
and the corresponding interface perturbation
\begin{eqnarray}
h_1(t) &=&\left [ h_1(0)/h_0(0) + f_{z1,2}^B \delta \gamma
  \sin t + f_{z1,3}^B \delta^2 \gamma^2 (\sin 2t) /4 \right. 
\nonumber\\
  &&\left. - (iq_y \lambda_2 + \delta f_{z1,4}^B) \gamma (1-\cos t) 
\right ] h_0(t),
\label{eq_h1}
\end{eqnarray}
with $h_0(t)$ given in Eq. (\ref{eq_h0}).
(It is convenient to choose $h_1(0)=0$ at time $t=0$ so that the
initial condition for $h$ is $h(t=0)=h_0(0)$, independent of $Re$.)

Equation (\ref{eq_sig1}) determines stability. The first term of the r.h.s. is
proportional to rescaled surface tension and is always negative, 
indicating the stabilizing effect of
surface tension. The second term, proportional to
$\gamma^2$, incorporates the effect of the imposed shear flow and
tends to destabilize the planar boundary. Detailed results are shown in the
next section.

\section{RESULTS}
\label{sec_results}

Given the results of the previous section, the stability of the boundary
depends on the two viscosity coefficients 
$\alpha_1$ and $\alpha_{56}$, on the domain thickness $d_A$,
the shear strain $\gamma$, as well as the
Reynolds number $Re$ and the rescaled surface tension $\Gamma'$ (both
$\omega$ dependent). Here we focus on three characteristic regimes of
$\Gamma'$ (all with $Re \ll 1$ as appropriate for typical copolymers): 
$\Gamma' = {\cal O}(1)$, $\Gamma'/Re = {\cal O}(1)$, and 
$\Gamma'/Re \ll 1$, corresponding to different ranges of shear frequencies.

\subsection{$Re \rightarrow 0$ and $\Gamma' = {\cal O}(1)$}

The first regime of interest is that of very small Reynolds number
($Re \rightarrow 0$) and finite surface tension ($\Gamma' = \Gamma_0
={\cal O}(1)$), which corresponds to very low frequency $\omega$
according to the analysis at the beginning of Sec. \ref{sec_analy}
\ref{subsec_sol} The Floquet exponent (or the perturbation growth
rate) is then well approximated by zeroth-order result $\sigma_0$ in
Eq. (\ref{eq_sig0}). Our calculations give $\sigma_0 \le 0$
(i.e., $f_{z0,1}^B(q_x,q_y) \le 0$) for
all wave numbers $q_x$ and $q_y$; thus, the system is always stable,
resulting in the coexistence of parallel and perpendicular domains
under shear flow. The stabilizing effect of
surface tension dominates, and $\sigma_0 \propto \Gamma_0$ as shown in
Eq. (\ref{eq_sig0}). 

\subsection{$Re \ll 1$, $\Gamma' \ll 1$, and $\Gamma'/Re
  = {\cal O}(1)$}
\label{subsec_O_Re}

In an intermediate range of $\omega$, $\Gamma'$ is of order ${\cal 
O}(Re)$ as discussed in Sec. \ref{sec_analy} \ref{subsec_sol} 
The perturbation growth rate can be written as
$\sigma=Re ~\sigma_1$ with the first-order exponent $\sigma_1$
given in Eq. (\ref{eq_sig1}). Note that
$f_{z1,1}^B=f_{z0,1}^B \leq 0$ whereas the
maximum value of $f_{z1,3}^B$ can be positive depending on system
parameters $\alpha_1$, $\alpha_{56}$, and $d_A$, leading to a
competition between the stabilizing effect of surface tension
(proportional to $\Gamma_1=\Gamma'/Re$) and the destabilizing
influence of the imposed
shear (proportional to $\gamma^2$). Note that $\Gamma_1$ can also be 
expressed from Eqs. (\ref{eq_G}) and (\ref{eq_Re}) as
\begin{equation}
\Gamma_1 = 1/We = [\Gamma/(\rho d^3)] \omega^{-2} \equiv \theta \omega^{-2},
\label{eq_G1}
\end{equation}
with $We$ the Weber number and $\theta = \Gamma/(\rho d^3)$. Thus, a
system would be more unstable for
larger shear strain $\gamma$ and frequency $\omega$.

To examine the onset of instability, we have carried out a numerical
evaluation of $\sigma$ for a range of typical system parameters. 
For a typical copolymer system, $\rho = 1$ g cm$^{-3}$, 
$\Gamma = 1$ dyne/cm, $\eta = 10^4$ P, and $d = 1$ cm, with Reynolds
number $Re = (10^{-4} {\rm s}) \omega$ and rescaled surface tension
$\Gamma' = (10^{-4} {\rm s}^{-1})/ \omega$. The two dimensionless
viscosities are set as: $\alpha_1=1$, and $\alpha_{56}=-0.9$
(corresponding to an effective viscosity $\mu_B=1/10$ and thus a ratio
$m=\mu_A/\mu_B=10$) or $\alpha_{56}=9$ (corresponding to $\mu_B=10$
and $m=1/10$).

\begin{figure}
\centerline{\epsfig{figure=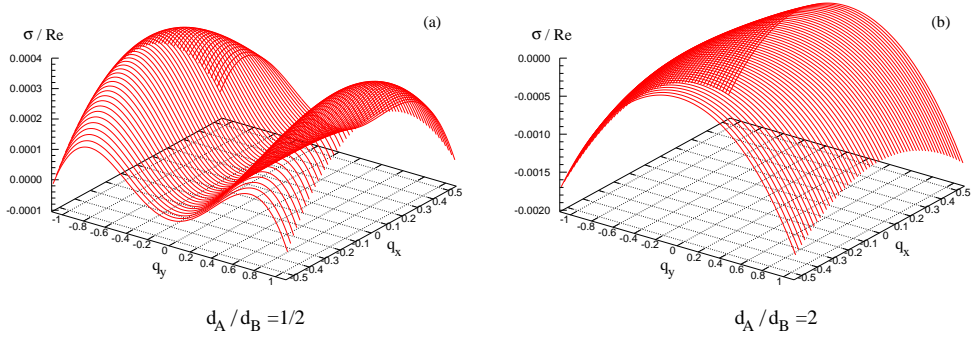,width=5.5in}}
\caption{Growth rate $\sigma /Re$ as a function of wave numbers $q_x$
  and $q_y$, for $\alpha_1=1$, $\alpha_{56}=-0.9$, $\gamma=1$,
  $Re=5 \times 10^{-4}$, as well as (a) $d_A=1/3$ (with $d_A/d_B=1/2$),
  with maximum growth rate $\sigma_{\rm max}=3.4 \times
  10^{-4} Re$ found at ${\bf q}^{\rm max}=(0,\pm 0.89)$, and (b)
  $d_A=2/3$ (with $d_A/d_B=2$), indicating that $\sigma \leq 0$ at all
  wavevectors.}
\label{fig_sigma}
\end{figure}

\begin{figure}
\centerline{\epsfig{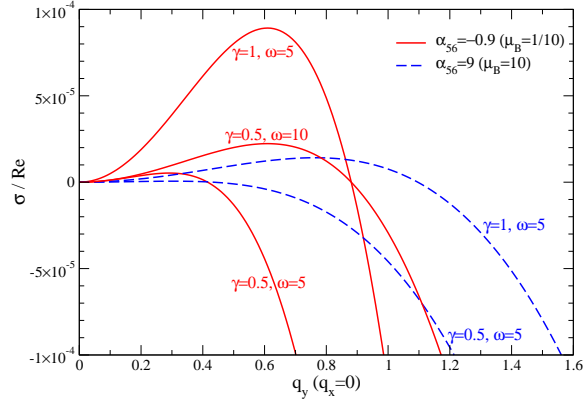}}
\caption{Growth rate $\sigma /Re$ versus $q_y$ at $q_x=0$ for
  $d_A/d_B=1$ and different values of $\gamma$, $\omega$, and
  $\alpha_{56}$.}
\label{fig_sigma2}
\end{figure}

Figure \ref{fig_sigma} shows the growth rate $\sigma_1=\sigma /Re$
as a function of wave numbers $(q_x,q_y)$, for $\alpha_{56}=-0.9$ and
two different domain thicknesses $d_A=1/3$ (with $d_A/d_B=1/2$) and $2/3$ 
(with $d_A/d_B=2$). The most dangerous
wave numbers are near $q_x = 0$, as can be seen in
Fig. \ref{fig_sigma}a. The figure shows that the maximum growth rate
($\sigma_{\rm max}=3.4 \times 10^{-4} Re$) occurs at $q_x^{\rm max}=0$
and $q_y^{\rm max}=\pm 0.89$. 
Results for the larger ratio $d_A/d_B=2$ are shown in
Fig. \ref{fig_sigma}b, indicating that the
system is stable. When $\alpha_{56}=9$, we obtain opposite stability
results: a small ratio $d_A/d_B=1/2$ corresponds to a stable configuration,
whereas instability occurs for $d_A/d_B=2$, with $\sigma_{\rm
 max}$ also found at $q_x = 0$. It is also interesting to
note that at $d_A/d_B=1$, instability occurs
for all values of $\alpha_{56}$ (i.e., for all effective
viscosity contrast). Figure \ref{fig_sigma2} shows the maximum growth
rate for $\gamma=1$ and 
$0.5$, $\omega=5$ and $10$ (s$^{-1}$) (corresponding to $Re=5 \times
10^{-4}$ and $10^{-3}$), and two different viscosities 
$\alpha_{56}=-0.9$ and $9$. As the shear strain 
amplitude $\gamma$ or frequency $\omega$ decreases, the range of
unstable wavenumbers also decreases. This observation motivates the
analysis of long wave solutions presented in subsection D.

\begin{figure}
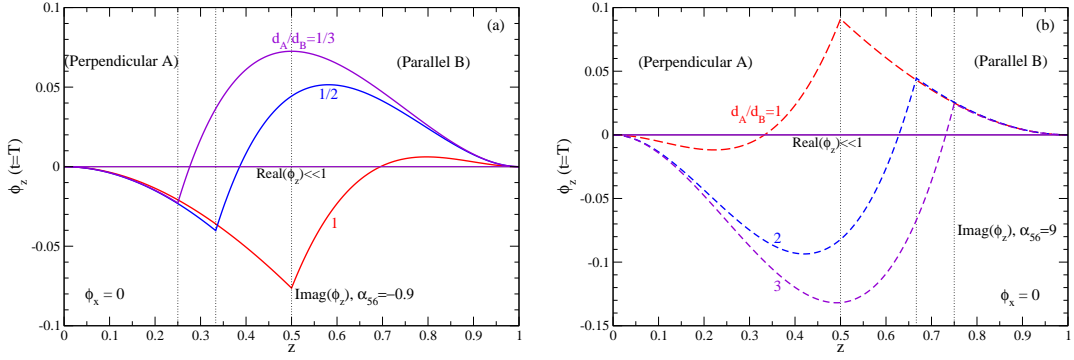

\centerline{\epsfig{figure=fig5a.eps,width=2.7in} \hskip 0.3cm
  \epsfig{figure=fig5b.eps,width=2.7in}}
\caption{Spatial dependence of the velocity amplitude $\phi_z$
  at time $t=T$ ($=2\pi$), at the
  most unstable wave vectors $(q_x^{\rm max}=0,q_y^{\rm max})$
  given by Figs. \ref{fig_sigma} and \ref{fig_sigma2}. All the
  curves here correspond to unstable configuration with $\sigma>0$, with
  parameters $Re = 5 \times 10^{-4}$, $\alpha_1=1$, and $\gamma=1$. 
  (a) $\alpha_{56}=-0.9$ ($\mu_B=1/10$), for $d_A/d_B=1$, $1/2$ and
  $1/3$; (b) $\alpha_{56}=9$ ($\mu_B=10$), for $d_A/d_B=1$, $2$, and 
  $3$. The locations of domain interface are indicated by dotted lines.}
\label{fig_phi_z}
\end{figure}

\begin{figure}
\centerline{\epsfig{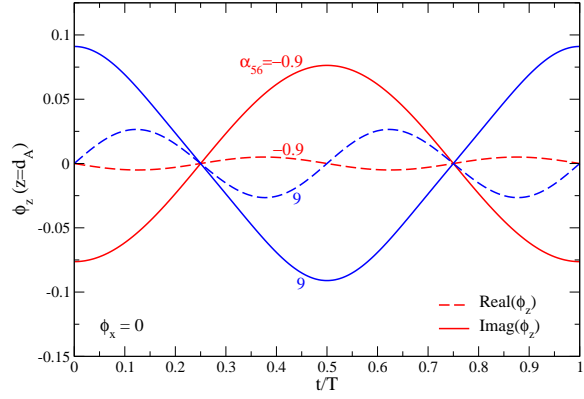}}
\caption{Temporal dependence of the velocity amplitude $\phi_z$ over a
  period $T$ at the interface $z=d_A$, with $d_A/d_B=1$, and other
  parameters the same as those of Fig. \ref{fig_phi_z}.}
\label{fig_phi_t}
\end{figure}

\begin{figure}
\centerline{\epsfig{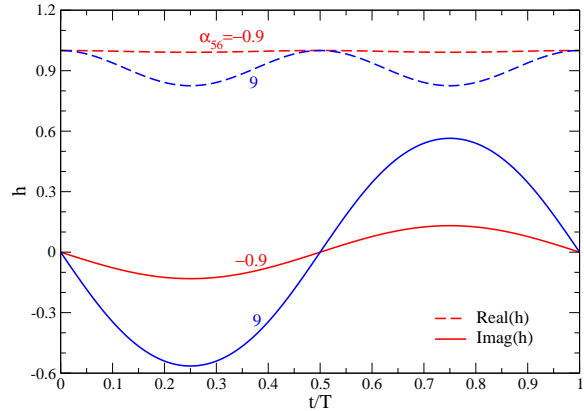}}
\caption{Temporal dependence of the interface perturbation $h$ 
  over a period $T$, with parameters the same as
  those of Fig. \ref{fig_phi_t}.}
\label{fig_h}
\end{figure}

The perturbed velocity fields are defined by Eq. (\ref{eq_floquet}).
The velocity associated with the most unstable wavevector $(q_x=0,
q_y \neq 0)$ is given by 
$$\hat{u}_x^{\rm max}=0 \quad {\rm and} \quad \hat{u}_z^{\rm max}=
\phi_z \exp(\sigma_{\rm max} t),$$ 
with $\phi_z$ determined by Eqs.
(\ref{eq_phi_z0B}), (\ref{eq_phi_z0A_qx0}), (\ref{eq_phi_z1B}), and
(\ref{eq_phi_z1A_qx0}). Results for $\phi_z$ at the most unstable
wavenumbers are shown in Fig. \ref{fig_phi_z} (at 
$t=T$), and in Fig. \ref{fig_phi_t} (at the interface $z=d_A$), 
for $\alpha_1=1$, $\gamma=1$, $Re=5 \times 10^{-4}$ (with 
$\omega=5$ s$^{-1}$), as well 
as for a variety of domain thickness ratios $d_A/d_B$ and different
viscosity coefficients $\alpha_{56}=-0.9$ and $9$. 
The corresponding interfacial perturbation
$h(t)$, defined in Eq. (\ref{eq_floquet}), is shown in Fig. \ref{fig_h}. 
Although the velocity fields near the interface are sensitive to the
viscosity contrast and thickness ratios (e.g, the temporal dependence of $\phi_z$
for $\mu_B=1/10$ (with $\alpha_{56}=-0.9$) and $\mu_B=10$ ($\alpha_{56}=9$) is
out of phase at the interface, as shown in Fig. \ref{fig_phi_t}), the qualitative
results are the same: The instability develops around the interface, 
and relaxes into the perpendicular (A) and parallel (B) bulk
regions. Importantly, the perturbation flow is directed along the $z$
direction  (and $y$, due to the incompressibility condition), and hence it is
transverse to the parallel lamellae. If the monomer density order parameter
were allowed to diffuse, this secondary flow would  result in the distortion
of parallel lamellae, but not perpendicular.

\subsection{$Re \ll 1$ and $\Gamma'/Re \ll 1$}

In the limit $\Gamma'/Re \ll 1$, which corresponds
to small surface tension $\Gamma$ or large enough $\omega$, the effect of
the surface tension is negligible at least for solutions up to first order.
This leads to $\Gamma_0=\Gamma_1=0$ in the solutions of
Sec. \ref{sec_analy} Thus $\sigma_0=0$, 
and according to Eq. (\ref{eq_sig1}) the first order growth
rate is
\begin{equation}
\sigma_1 = \frac{1}{2} \delta^2 \gamma^2 f_{z1,3}^B(q_x,q_y).
\label{eq_sig1_1}
\end{equation}
The function $f_{z1,3}^B(q_x,q_y)$ can be positive, and note that it
is independent of shear parameters $\gamma$ and $\omega$. The growth
rate is proportional to $\gamma^2$. 

\begin{figure}
\vskip 24pt
\centerline{\epsfig{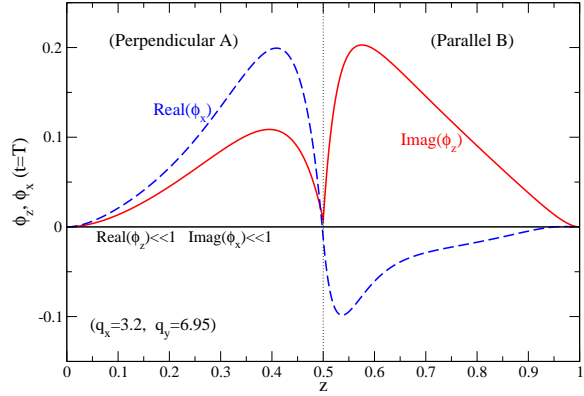}}
\caption{Velocity amplitudes $\phi_x$ and $\phi_z$ (at time $t=T$) as
  a function of position $z$, at the most unstable wave numbers $(q_x^{\rm
  max},q_y^{\rm max}) = (3.2,6.95)$. The parameters are chosen as
  $\alpha_{56}=-0.9$, $d_A/d_B=1$, $\gamma=1$, and $Re=10^{-2}$
  (corresponding to large $\omega=100$ s$^{-1}$).}
\label{fig_phi_w100}
\end{figure}

A typical profile for the velocity functions $\phi_x$ and $\phi_z$ is
shown in Fig. \ref{fig_phi_w100}. In contrast with the regime discussed in 
subsection B ($\Gamma'/Re = \Gamma_1 = {\cal O}(1)$),
the most unstable wave numbers $q_x^{\rm max}$ and $q_y^{\rm max}$ here
are both nonzero. As shown in Fig. \ref{fig_phi_w100} and unlike the
limit discussed in subsection B, perturbed
velocity fields develop along both $x$ and $z$ directions (i.e.,
$\phi_x, \phi_z \neq 0$), leading to
the distortion of both parallel and perpendicular regions. Therefore, 
although the parallel/perpendicular interface will move in response to the
hydrodynamic instability and the resulting secondary flows, 
the direction of motion and hence the
dominant lamellar orientation cannot be deduced from this analysis.

\subsection{Long wave solutions}
\label{subsec_longwave}

\begin{figure}
\centerline{\epsfig{figure=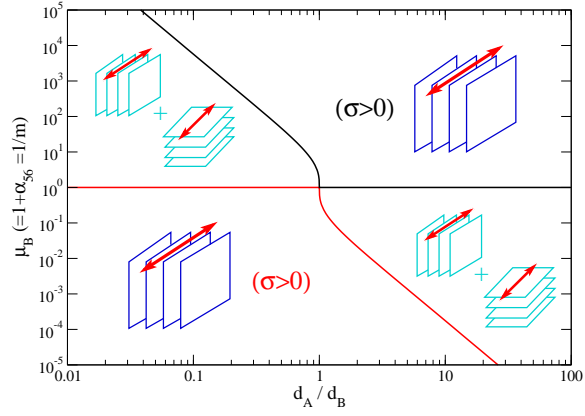,width=3.in}}
\caption{Stability diagram of thickness ratio $d_A/d_B$ versus
  viscosity contrast $\mu_B$ ($=m^{-1}$).
  For the unstable regime with $\sigma>0$, the perpendicular
  phase is selected over the parallel one, while the stability leads to
  the coexistence of parallel and perpendicular orientations.}
\label{fig_d_mu}
\end{figure}

The calculations presented so far show that in the limit 
$Re \ll 1$, $\Gamma' \ll 1$, with $\Gamma'/Re = {\cal O}(1)$,
the wave numbers associated with
instability lie at $q_x=0$ and small $q_y$, as seen in
Fig. \ref{fig_sigma}. Thus, we discuss next a long wave approximation  
to the stability analysis. We expand the solutions of
Sec. \ref{sec_analy}C in powers of $q$ ($=q_y$ here), and find that
\begin{equation}
f_{z1,1}^B = f_{z0,1}^B = - f_0\ q^4 + {\cal O}(q^6),
\label{eq_fz1_0}
\end{equation}
with
\begin{equation}
f_0 = \frac{1}{3 \Delta} d_A^3 d_B^3 (1+\alpha_{56} d_A) >0,
\label{eq_f0}
\end{equation}
and
\begin{equation}
f_{z1,3}^B = f_1\ q^2 + f_2\ q^4  + {\cal O}(q^6),
\label{eq_fz1_3}
\end{equation}
where
\begin{equation}
f_1 = \frac{1}{60 \Delta^3} d_A^2 d_B^2 \alpha_{56} f_{11} f_{12},
\label{eq_f1}
\end{equation}
with
\begin{eqnarray}
f_{11} &=& (1+\alpha_{56} d_A^2)^2 + 4\alpha_{56} d_A^2 d_B 
(1+\alpha_{56} d_A) >0 \nonumber\\
f_{12} &=& (d_A-d_B)(d_A^2 + d_B^2) + d_A^8 \alpha_{56}^4 + 
2d_A^5 [d_A d_B^2 + 2(d_A-d_B)] \alpha_{56}^3 \nonumber\\
&+& 2d_A^2 [ 3(d_A-d_B)^2 + (1-d_A^2)^2-2d_B^3 (1+d_B)^2]\alpha_{56}^2 \nonumber\\
&+& 2d_A [ 2(d_A-d_B)^2+3d_A d_B^2 - 4(1+d_A)d_B^4] \alpha_{56}, 
\label{eq_f11_12}
\end{eqnarray}
and $f_2$ a complicated but known function of $d_A$, $\alpha_1$,
and $\alpha_{56}$. Here $\Delta$ is defined as
\begin{equation}
\Delta = (d_B^2-\mu_B d_A^2)^2+4\mu_B d_A d_B >0.
\label{eq_Delta}
\end{equation}
Therefore, the first-order Floquet exponent (\ref{eq_sig1}) can be
rewritten as 
\begin{eqnarray}
\sigma_1 &=& \theta f_{z1,1}^B \omega^{-2} + \frac{1}{2} \delta^2
f_{z1,3}^B \gamma^2 \nonumber\\
&=& \frac{1}{2} \delta^2 f_1 \gamma^2 q^2 - (\theta f_0 \omega^{-2} 
- \frac{1}{2} \delta^2 f_2 \gamma^2 ) q^4.
\label{eq_sigma1}
\end{eqnarray}
Thus, when $f_1>0$ at small $q$ we have $\sigma_1>0$ for all $\gamma$
and $\omega$. From the definition of $f_1$, we note that stability is
determined by the sign of $\alpha_{56}
f_{12}$ (Eqs. (\ref{eq_f1}) and (\ref{eq_f11_12})) which is itself
a function of $\alpha_{56}$ and $d_A$ only, and independent of shear
parameters $\gamma$ and $\omega$. The calculated stability diagram of
$d_A/d_B$ vs. $\mu_B$ ($=1+\alpha_{56}$) is shown in
Fig. \ref{fig_d_mu}. The diagram is symmetric with respect to $d_A/d_B
\rightarrow (d_A/d_B)^{-1}$ and $\mu_B \rightarrow \mu_B^{-1}$. (Note
that at $d_A/d_B=1$ (i.e., $d_A=1/2$) instability is found
for all values of $\mu_B$, in agreement with the numerical results in
subsection \ref{subsec_O_Re}) This diagram reveals the analog of the
thin layer effect in the interfacial instability caused by
viscosity stratification of two superposed Newtonian fluids
[\citet{re:hooper85}]. When the thinner domain has the larger
effective viscosity, instability occurs. Unlike the
Newtonian case, here the effective viscosity contrast is caused by the 
orientation dependence of the dissipative part of the stress tensor.
Note the for a polycrystalline sample, although the
value of $\alpha_{56}$ ($\mu_B$) would be determined by the specific
copolymer considered, the thickness ratio $d_A/d_B$
would vary from domain to domain. Thus, according to 
Fig. \ref{fig_d_mu}, the development of the instability would differ in
different portions of a large sample. 

The largest perturbation growth rate can be obtained from
Eq. (\ref{eq_sigma1}). When $\theta f_0 \omega^{-2} - \frac{1}{2}
\delta^2 f_2 \gamma^2 >0$, or
\begin{equation}
\gamma^2 \omega^2 < \frac{2\theta f_0}{\delta^2 f_2},
\label{eq_gw2}
\end{equation}
we obtain
\begin{eqnarray}
\sigma_1^{\rm max} &=& \frac{\delta^4 f_1^2}{16 (\theta f_0 -
  \delta^2 f_2 \gamma^2 \omega^2 /2)} \gamma^4 \omega^2, \nonumber\\
q_{\rm max} &=& \frac{1}{2} \delta \left (\frac{f_1}{\theta f_0 -
  \delta^2 f_2 \gamma^2 \omega^2 /2} \right )^{1/2} \gamma \omega.
\label{eq_sig_q}
\end{eqnarray}
These formulae show that both maximum growth rate and most unstable
wavenumber increase with shear amplitude $\gamma$ and frequency $\omega$,
in agreement with the numerical results of subsection B and 
Fig. \ref{fig_sigma2}.

\section{DISCUSSION}
\label{sec_discuss}

The analysis given is purely of hydrodynamic nature and makes no
reference to the response of the lamellar phases to the flows
considered. The fully coupled problem is very complex, but the flow
analysis conducted here can be used to argue indirectly about orientation selection.
The flow perturbation ${\bf u}$ will advect the lamellae through 
the advection term ${\bf u} \cdot {\bm \nabla}\psi$ in
Eq. (\ref{eq_G-L}), with $\psi$ the monomer concentration. Parallel
lamellae are marginal to velocity fields along the
$x$ and $y$ directions since these flows are parallel to the planes of
constant $\psi$, but will be distorted by flows in the
$z$ direction (Fig. \ref{fig_conf}). Conversely, 
perpendicular lamellae are unaffected by flows along either $z$ or $y$,
but distorted by those along the $x$ direction. The instability
mode given in Sec. \ref{sec_results}B for $\Gamma'/Re =
{\cal O}(1)$ which might be of most experimental relevance (if we
estimate the order of surface tension $\Gamma$ from the
polymer-polymer interfacial energy [\citet{re:helfand72,re:gido94}], 
as discussed in Sec. \ref{sec_analy}C) and hence is our
focus here, is associated with secondary flows with $u_x=0$ and
$u_z \neq 0$; thus parallel lamellae are compressed or expanded, while
the lamellar configuration in the perpendicular 
region remains unaffected due to the absence of modulation along its normal.
Distortion of parallel lamellae would create a relative imbalance of free
energy ${\cal F}$ in the two domains: ${\cal F}_{\rm Parallel} >
{\cal F}_{\rm Perpendicular}$. This free energy imbalance would be
relieved through the motion of the domain boundary towards the distorted
parallel region. Therefore we would anticipate that a consequence of
the shear flow and the resulting interfacial instabilities would be
the growth of the perpendicular region at the expense of the parallel one.

Therefore, the stability diagram of Fig. \ref{fig_d_mu} can be used to
indirectly address orientation selection, suggesting coexistence of
parallel and perpendicular domains (in the hydrodynamically stable
regime) or the selection of the perpendicular 
orientation (in the unstable regime). It would be interesting to
examine an experimental system composed of only two domains of
parallel and perpendicular orientations, for verifying our predictions
such as the change of instability with domain thickness ratio
$d_A/d_B$ and viscosity contrast $\mu_B$, measuring the
viscosity contrast from the location of the instability boundary, and 
further studying the domain evolution beyond the instability stage.

Polycrystalline samples of the type present in all shear aligning
experiments involve a distribution of grain sizes and orientations. We
address next how the results just obtained may provide a criterion for
orientation selection under certain circumstances. Experiments 
[\citet{re:gido93,re:qiao00}] reveal the presence of grain
boundaries separating domains of different orientations, and we argue
that the motion of these grain boundaries under the imposed shear
affects the selection process. This mechanism is different than other
suggestions in the literature involving grain rotation,
domain instabilities, or other effects of microscopic
origin that are related to block architecture (looping and bridging)
[\citet{re:wu04,re:wu05}].

Earlier research has shown that boundaries of domains with a lamellar
normal that has a component along the transverse direction will move,
leading to a decrease in the size of the domain. The shear increases
the free energy density of the transverse domain and originates
diffusive monomer redistribution at the boundary to reduce the extent
of the phase of higher free energy
[\citet{re:huang03,re:huang04}]. Since the free energy of neither
parallel nor perpendicular lamellae is affected by the shear (at least
in the low frequency range of $\omega \ll \omega_c$ in which the
Leibler [\citet{re:leibler80}] or Ohta-Kawasaki [\citet{re:ohta86}]
free energies are a good approximation), one can generally expect
grain boundaries to move toward the transverse phase. Our interest in
this paper is therefore in possible physical mechanisms that would
account for the motion of boundaries separating domains of parallel
and perpendicular orientations. According to Fig. \ref{fig_d_mu}, if
$\alpha_{56} >0$ (as might be appropriate, for example, for PEP-PEE
diblocks) an initially large perpendicular domain (A) adjacent to a
smaller parallel domain (B) (so that $d_A/d_B$ is large) would grow
even larger. Although our analysis does not hold beyond the linear
stage of boundary deformation, it seems unlikely that any nonlinearity
could saturate boundary distortion and lead to a stationary but
corrugated boundary. Therefore we would predict that the perpendicular
orientation will be selected for $\alpha_{56} >0$. If, on the other
hand, $\alpha_{56} <0$ (as would be appropriate, for example, for
PS-PI diblocks), the situation is more complicated. Instability now
occurs for $d_A/d_B$ small, leading to growth of the perpendicular
domain and hence to an increase of the characteristic scale $d_A$. To
the extent that, in a sufficiently large system the boundary remains
quasi-planar, the stability boundary in Fig. \ref{fig_d_mu} would be
reached. Once inside the stable region, any remaining curved
boundaries would be expected to relax to planarity (driven by excess
free energy reduction), as the planar boundary would no longer be
unstable under shear. Therefore, in the case of $\alpha_{56}<0$, we
would anticipate coexistence of parallel and perpendicular domains, or
perhaps a dependence of the selected orientation on initial condition
or sample history.

\begin{figure}
\centerline{\epsfig{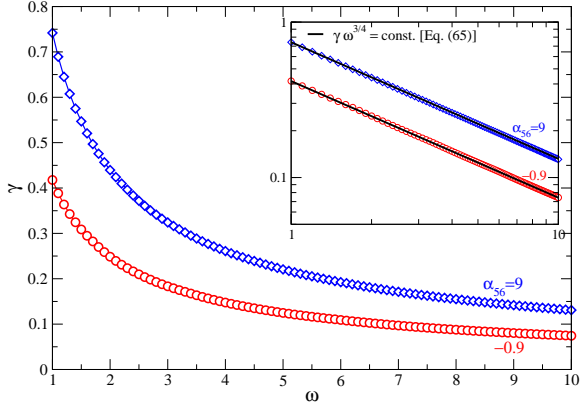}}
\caption{Lines of constant $\sigma_{\rm max}$ ($=10^{-11}$) for 
  $d_A/d_B=1$. Symbols (circles and diamonds) are from
  numerical calculations, while the solid lines in the inset follow
  the scaling given by the long wave approximation in
  Eq. (\ref{eq_sig_max}), that is, $\gamma \omega^{3/4}=const.$}
\label{fig_gw}
\end{figure}

We finally examine the dependence of the largest growth rate
$\sigma_{\rm max}$ on the shear amplitude $\gamma$ and angular
frequency $\omega$ near the onset of instability. Note that the
stability boundaries of Fig. \ref{fig_d_mu} are independent of both
parameters, but near onset where $\sigma \sim {\cal O}(Re) \ll 1$,
experiments might detect an effective stability boundary located at
the point on which $1/\sigma_{\rm max}$ is of the order of the
observation time of the experiment. From Eq. (\ref{eq_Re})
we have
$$
\sigma_1^{\rm max}=\sigma_{\rm max}/Re 
= \sigma_{\rm max} \eta/(\rho d^2 \omega),
$$
and then given Eq. (\ref{eq_sig_q}) we find that
\begin{equation}
\gamma^2 \omega^2 \left ( \gamma^2 \omega + \frac{8f_2 \eta \sigma_{\rm max}}
{\delta^2 f_1^2 \rho d^2} \right ) 
= \frac{16\theta f_0 \eta \sigma_{\rm max}}{\delta^4
  f_1^2 \rho d^2}, \label{eq_gw1}
\end{equation}
with associated wavenumber
\begin{equation}
q_{\rm max} = \left ( \frac{\eta \sigma_{\rm max}}{\theta f_0 \rho
 d^2} \right )^{1/4} \omega^{1/4}. \label{eq_qmax_b}
\end{equation}
Usually $\gamma^2 \omega \gg 8|f_2| \eta \sigma_{\rm max}/ 
(\delta^2 f_1^2 \rho d^2)$ for very small value of $\sigma_{\rm max}$,
and thus we obtain 
\begin{equation}
\sigma_{\rm max} = \frac{\rho d^2 \delta^4 f_1^2}{16 \theta f_0 \eta}
\gamma^4 \omega^3,
\label{eq_sig_max}
\end{equation}
which scales as $\gamma \omega^{3/4}$ for a given block
copolymer. Therefore, the line of constant $\sigma_{\rm max}$ is given
by $\gamma \omega^{3/4}=const.$ This result is consistent with our
numerical evaluation of solutions in Sec. \ref{sec_analy}C.2, as shown
in Fig. \ref{fig_gw} for two cases of $\alpha_{56}=-0.9$ (circles) and
$9$ (diamonds). Note that the data in the log-log plot of the
inset is well fitted to a straight line with slope $-3/4$. For small
enough $\sigma_{\rm max}$ (e.g., $=10^{-11}$ as in Fig. \ref{fig_gw}), 
which is the order of the inverse
experimental observation time, this line can correspond to an
effective stability boundary above which the instability becomes
experimentally observable. 

To our knowledge, the only experimental determination in
$\gamma$--$\omega$ space of the regions in which 
parallel or perpendicular lamellae are selected has been carried out
for PS-PI block copolymers [\citet{re:maring97,re:leist99}]. It has
been found that the line separating regions of perpendicular and
parallel orientations was approximately given by
$\gamma \omega =const.$ No results for
copolymers with $\alpha_{56}>0$ (such as PEP-PEE) are available. Given
that our predictions of effective boundary only apply to this case, it
would be of interest to repeat the experiments for this type of
copolymers.

The above discussion corresponds to a range of
shear frequency so that $\Gamma'/Re = {\cal O}(1)$. For very low
frequencies so that $Re \rightarrow 0$ while $\Gamma' = {\cal O}(1)$, 
a parallel/perpendicular configuration is always
stable due to the dominant effect of surface tension (as given in
Sec. \ref{sec_results}A). 

In combination with the stability of the lowest frequencies, our
results in Fig. \ref{fig_gw} indicate that the perpendicular
orientation would be observed for large enough $\omega$ 
and $\gamma$. Otherwise, both parallel and perpendicular lamellae
would coexist, and the selection between them would depend on experimental
details such as quenched or annealed history of the sample and the
starting time of the shear, as found in PS-PI copolymer samples 
[\citet{re:patel95,re:maring97,re:larson99}] and cannot be addressed
by the stability analysis here.

\section{SUMMARY}

The assumption of a dissipative part of the stress tensor
$\sigma_{ij}^D$ which is compatible with the uniaxial symmetry of
a lamellar phase leads to an effective dynamic viscosity that depends
on the orientation of the lamellae relative to the shear. We expect
this functional form of $\sigma_{ij}^D$ to capture the low
frequency and long wavelength response to the lamellar phase, without
making reference to the microscopic origin of the viscosity
coefficients. We have explored here the consequence of this assumption
on a configuration comprising parallel and perpendicular
domains. Experimental evidence suggests that these two orientations
are prevalent in shear aligning experiments, and we believe that the
type of rheology proposed here may contribute to our understanding of
the orientation selected as a function of the parameters of the block
and of the shear.

In particular, we have shown that an oscillatory shear imposed on a
block copolymer configuration comprising lamellar domains of parallel
and perpendicular orientations can cause instability at the domain
interface. The instability manifests itself
by finite wavenumber undulations of the velocity field along the
direction normal to parallel lamellae, which we argue would ultimately
result in the growth of the perpendicular region at the expense of the
parallel one. Our results indicate that the instability,
and the selection of the perpendicular orientation, occur at an
intermediate frequency range of small but finite value
of $Re$, and depend on both viscosity contrast and
domain thickness ratio. This instability is analogous to the thin layer
effect in stratified fluids; that is, the system is unstable when
the thinner domain is more viscous. On the
other hand, at very low frequencies ($Re \rightarrow 0$),
coexistence of parallel and perpendicular lamellae is found as implied by
hydrodynamic stability. Also, in contrast to
previous studies, the selection mechanism between
parallel and perpendicular orientations introduced here is of
dynamical nature, and an indirect consequence of 
the secondary flows generated by the hydrodynamic instability
of the two-domain interface. It would be interesting to test our
predictions experimentally in a test configuration of block copolymers
as we have discussed here. Note that the frequency range
studied here is $\omega<\omega_c$ in which polymer chains remain
relaxed and hence the details of the individual blocks are not
important. Thus our results should be independent of number and type of
blocks in a copolymer.

\section*{ACKNOWLEDGMENTS}

This work has been supported by the National Science Foundation
under grant DMR-0100903, and by NSERC Canada.

\newpage
\begin{center}
FIGURE CAPTIONS
\end{center}

\begin{itemize}
\item[FIG. 1.] Three lamellar orientations (Parallel, Perpendicular, and
  Transverse) under shear flow.

\item[FIG. 2.] A parallel/perpendicular configuration subjected to
  oscillatory shear flow.

\item[FIG. 3.] Growth rate $\sigma /Re$ as a function of wave numbers $q_x$
  and $q_y$, for $\alpha_1=1$, $\alpha_{56}=-0.9$, $\gamma=1$,
  $Re=5 \times 10^{-4}$, as well as (a) $d_A=1/3$ (with $d_A/d_B=1/2$),
  with maximum growth rate $\sigma_{\rm max}=3.4 \times
  10^{-4} Re$ found at ${\bf q}^{\rm max}=(0,\pm 0.89)$, and (b)
  $d_A=2/3$ (with $d_A/d_B=2$), indicating that $\sigma \leq 0$ at all
  wavevectors.

\item[FIG. 4.] Growth rate $\sigma /Re$ versus $q_y$ at $q_x=0$ for
  $d_A/d_B=1$ and different values of $\gamma$, $\omega$, and
  $\alpha_{56}$.

\item[FIG. 5.] Spatial dependence of the velocity amplitude $\phi_z$
  at time $t=T$ ($=2\pi$), at the
  most unstable wave vectors $(q_x^{\rm max}=0,q_y^{\rm max})$
  given by Figs. \ref{fig_sigma} and \ref{fig_sigma2}. All the
  curves here correspond to unstable configuration with $\sigma>0$, with
  parameters $Re = 5 \times 10^{-4}$, $\alpha_1=1$, and $\gamma=1$. 
  (a) $\alpha_{56}=-0.9$ ($\mu_B=1/10$), for $d_A/d_B=1$, $1/2$ and
  $1/3$; (b) $\alpha_{56}=9$ ($\mu_B=10$), for $d_A/d_B=1$, $2$, and 
  $3$. The locations of domain interface are indicated by dotted lines.

\item[FIG. 6.] Temporal dependence of the velocity amplitude $\phi_z$ over a
  period $T$ at the interface $z=d_A$, with $d_A/d_B=1$, and other
  parameters the same as those of Fig. \ref{fig_phi_z}.

\item[FIG. 7.] Temporal dependence of the interface perturbation $h$ 
  over a period $T$, with parameters the same as
  those of Fig. \ref{fig_phi_t}.

\item[FIG. 8.] Velocity amplitudes $\phi_x$ and $\phi_z$ (at time $t=T$) as
  a function of position $z$, at the most unstable wave numbers $(q_x^{\rm
  max},q_y^{\rm max}) = (3.2,6.95)$. The parameters are chosen as
  $\alpha_{56}=-0.9$, $d_A/d_B=1$, $\gamma=1$, and $Re=10^{-2}$
  (corresponding to large $\omega=100$ s$^{-1}$).

\item[FIG. 9.] Stability diagram of thickness ratio $d_A/d_B$ versus
  viscosity contrast $\mu_B$ ($=m^{-1}$).
  For the unstable regime with $\sigma>0$, the perpendicular
  phase is selected over the parallel one, while the stability leads to
  the coexistence of parallel and perpendicular orientations.

\item[FIG. 10.] Lines of constant $\sigma_{\rm max}$ ($=10^{-11}$) for 
  $d_A/d_B=1$. Symbols (circles and diamonds) are from
  numerical calculations, while the solid lines in the inset follow
  the scaling given by the long wave approximation in
  Eq. (\ref{eq_sig_max}), that is, $\gamma \omega^{3/4}=const.$

\end{itemize}


\begin{thebibliography}{}
\newcommand{\enquote}[1]{``#1''}

\bibitem[Cates and Milner(1989)Cates and Milner]{re:cates89}
Cates, M.~E. and S.~T. Milner, \enquote{Role of shear in the
  isotropic-to-lamellar transition,} Phys. Rev. Lett. \textbf{62}, 1856 (1989).

\bibitem[Chen and {Vi\~nals}(2002)Chen and {Vi\~nals}]{re:chen02}
Chen, P. and J.~{Vi\~nals}, \enquote{Lamellar phase stability in diblock
  copolymers under oscillatory shear flows,} Macromolecules \textbf{35}, 4183
  (2002).

\bibitem[Chen and Kornfield(1998)Chen and Kornfield]{re:chen98b}
Chen, Z.~R. and J.~A. Kornfield, \enquote{Flow-induced alignment of lamellar
  block copolymer melts,} Polymer \textbf{39}, 4679 (1998).

\bibitem[de~Gennes and Prost(1993)de~Gennes and Prost]{re:degennes93}
de~Gennes, P.~G. and J.~Prost, \emph{The Physics of Liquid Crystals},
  Clarendon, Oxford (1993).

\bibitem[Drolet \emph{et~al.}(1999)Drolet, Chen and {Vi\~nals}]{re:drolet99}
Drolet, F., P.~Chen and J.~{Vi\~nals}, \enquote{Lamellae alignment by shear
  flow in a model of a diblock copolymer,} Macromolecules \textbf{32}, 8603
  (1999).

\bibitem[Ericksen(1960)Ericksen]{re:ericksen60}
Ericksen, J.~L., \enquote{Anisotropic fluids,} Arch. Ration. Mech. Anal.
  \textbf{4}, 231 (1960).

\bibitem[Forster \emph{et~al.}(1971)Forster, Lubensky, Martin, Swift and
  Pershan]{re:forster71}
Forster, D., T.~C. Lubensky, P.~C. Martin, J.~Swift and P.~S. Pershan,
  \enquote{Hydrodynamics of liquid crystals,} Phys. Rev. Lett. \textbf{26},
  1016 (1971).

\bibitem[Fredrickson(1994)Fredrickson]{re:fredrickson94}
Fredrickson, G.~H., \enquote{Steady shear alignment of block copolymers near
  the isotropic-lamellar transition,} J. Rheol. \textbf{38}, 1045 (1994).

\bibitem[Fredrickson and Bates(1996)Fredrickson and Bates]{re:fredrickson96}
Fredrickson, G.~H. and F.~S. Bates, \enquote{Dynamics of block copolymers,}
  Annu. Rev. Mater. Sci. \textbf{26}, 501 (1996).

\bibitem[Fredrickson and Helfand(1987)Fredrickson and
  Helfand]{re:fredrickson87}
Fredrickson, G.~H. and E.~Helfand, \enquote{Fluctuation effects in the theory
  of microphase separation in block copolymers,} J. Chem. Phys. \textbf{87},
  697 (1987).

\bibitem[Gido \emph{et~al.}(1993)Gido, Gunther, Thomas and Hoffman]{re:gido93}
Gido, S.~P., J.~Gunther, E.~L. Thomas and D.~Hoffman, \enquote{Lamellar diblock
  copolymer grain boundary morphology. 1. Twist boundary characterization,}
  Macromolecules \textbf{26}, 4506 (1993).

\bibitem[Gido and Thomas(1994)Gido and Thomas]{re:gido94}
Gido, S.~P. and E.~L. Thomas, \enquote{Lamellar diblock copolymer grain
  boundary morphology. 2. Scherk twist boundary energy calculations,}
  Macromolecules \textbf{27}, 849 (1994).

\bibitem[Goulian and Milner(1995)Goulian and Milner]{re:goulian95}
Goulian, M. and S.~T. Milner, \enquote{Shear alignment and instability of
  smectic phases,} Phys. Rev. Lett. \textbf{74}, 1775 (1995).

\bibitem[Guo(2006)Guo]{re:guo06}
Guo, H.~X., \enquote{Shear-induced parallel-to-perpendicular orientation
  transition in the amphiphilic lamellar phase: A nonequilibrium
  molecular-dynamics simulation study,} J. Chem. Phys. \textbf{124}, 054902
  (2006).

\bibitem[Gupta \emph{et~al.}(1995)Gupta, Krishnamoorti, Kornfield and
  Smith]{re:gupta95}
Gupta, V.~K., R.~Krishnamoorti, J.~A. Kornfield and S.~D. Smith,
  \enquote{Evolution of microstructure during shear alignment in a
  polystyrene-polyisoprene lamellar diblock copolymer,} Macromolecules
  \textbf{28}, 4464 (1995).

\bibitem[Helfand and Tagami(1972)Helfand and Tagami]{re:helfand72}
Helfand, E. and Y.~Tagami, \enquote{Theory of the interface between immiscible
  polymers. ii,} J. Chem. Phys. \textbf{56}, 3592 (1972).

\bibitem[Hooper(1985)Hooper]{re:hooper85}
Hooper, A.~P., \enquote{Long-wave instability at the interface between two
  viscous fluids: thin layer effects,} Phys. Fluids \textbf{28}, 1613 (1985).

\bibitem[Huang \emph{et~al.}(2003)Huang, Drolet and {Vi\~nals}]{re:huang03}
Huang, Z.-F., F.~Drolet and J.~{Vi\~nals}, \enquote{Motion of a
  transverse/parallel grain boundary in a block copolymer under oscillatory
  shear flow,} Macromolecules \textbf{36}, 9622 (2003).

\bibitem[Huang and {Vi\~nals}(2004)Huang and {Vi\~nals}]{re:huang04}
Huang, Z.-F. and J.~{Vi\~nals}, \enquote{Shear induced grain boundary motion
  for lamellar phases in the weakly nonlinear regime,} Phys. Rev. E
  \textbf{69}, 041504 (2004).

\bibitem[King \emph{et~al.}(1999)King, Leighton and McCready]{re:king99}
King, M.~R., D.~T. Leighton, Jr. and M.~J. McCready, \enquote{Stability of
  oscillatory two-phase couette flow: theory and experiment,} Phys. Fluids
  \textbf{11}, 833 (1999).

\bibitem[Koppi \emph{et~al.}(1992)Koppi, Tirrell, Bates, Almdal and
  Colby]{re:koppi92}
Koppi, K.~A., M.~Tirrell, F.~S. Bates, K.~Almdal and R.~H. Colby,
  \enquote{Lamellar orientation in dynamically sheared diblock copolymer
  melts,} J. Phys. II (France) \textbf{2}, 1941 (1992).

\bibitem[Larson(1999)Larson]{re:larson99}
Larson, R.~G., \emph{The Structure and Rheology of Complex Fluids}, Oxford
  University Press, New York (1999).

\bibitem[Leibler(1980)Leibler]{re:leibler80}
Leibler, L., \enquote{Theory of microphase separation in block copolymers,}
  Macromolecules \textbf{13}, 1602 (1980).

\bibitem[Leist \emph{et~al.}(1999)Leist, Maring, Thurn-Albrecht and
  Wiesner]{re:leist99}
Leist, H., D.~Maring, T.~Thurn-Albrecht and U.~Wiesner, \enquote{Double flip of
  orientation for a lamellar diblock copolymer under shear,} J. Chem. Phys.
  \textbf{110}, 8225 (1999).

\bibitem[Leslie(1966)Leslie]{re:leslie66}
Leslie, F.~M., \enquote{Some constitutive equations for anisotropic fluids,}
  Quart. J. Mech. Appl. Math. \textbf{19}, 357 (1966).

\bibitem[Maring and Wiesner(1997)Maring and Wiesner]{re:maring97}
Maring, D. and U.~Wiesner, \enquote{Threshold strain value for perpendicular
  orientation in dynamically sheared diblock copolymers,} Macromolecules
  \textbf{30}, 660 (1997).

\bibitem[Martin \emph{et~al.}(1972)Martin, Parodi and Pershan]{re:martin72}
Martin, P.~C., O.~Parodi and P.~S. Pershan, \enquote{Unified hydrodynamic
  theory for crystals, liquid crystals, and normal fluids,} Phys. Rev. A
  \textbf{6}, 2401 (1972).

\bibitem[Ohta and Kawasaki(1986)Ohta and Kawasaki]{re:ohta86}
Ohta, T. and K.~Kawasaki, \enquote{Equilibrium morphology of block copolymer
  melts,} Macromolecules \textbf{19}, 2621 (1986).

\bibitem[Patel \emph{et~al.}(1995)Patel, Larson, Winey and
  Watanabe]{re:patel95}
Patel, S.~S., R.~G. Larson, K.~I. Winey and H.~Watanabe, \enquote{Shear
  orientation and rheology of a lamellar polystyrene-polyisoprene block
  copolymer,} Macromolecules \textbf{28}, 4313 (1995).

\bibitem[Pinheiro \emph{et~al.}(1996)Pinheiro, Hajduk, Gruner and
  Winey]{re:pinheiro96}
Pinheiro, B.~S., D.~A. Hajduk, S.~M. Gruner and K.~I. Winey,
  \enquote{Shear-stabilized bi-axial texture and lamellar contraction in both
  diblock copolymer and diblock copolymer/homopolymer blends,} Macromolecules
  \textbf{29}, 1482 (1996).

\bibitem[Pinheiro and Winey(1998)Pinheiro and Winey]{re:pinheiro98}
Pinheiro, B.~S. and K.~I. Winey, \enquote{Mixed parallel-perpendicular
  morphologies in diblock copolymer systems correlated to the linear
  viscoelastic properties of the parallel and perpendicular morphologies,}
  Macromolecules \textbf{31}, 4447 (1998).

\bibitem[Qiao and Winey(2000)Qiao and Winey]{re:qiao00}
Qiao, L. and K.~I. Winey, \enquote{Evoluton of kink bands and tilt boundaries
  in block copolymers at large shear strains,} Macromolecules \textbf{33}, 851
  (2000).

\bibitem[Soddemann \emph{et~al.}(2004)Soddemann, Auernhammer, Guo, {D\"{u}nweg}
  and Kremer]{re:soddemann04}
Soddemann, T., G.~K. Auernhammer, H.~Guo, B.~{D\"{u}nweg} and K.~Kremer,
  \enquote{Shear-induced undulation of smectic-{A}: Molecular dynamics
  simulations vs. analytical theory,} Eur. Phys. J. E \textbf{13}, 141 (2004).

\bibitem[Wu \emph{et~al.}(2004)Wu, Lodge and Bates]{re:wu04}
Wu, L., T.~P. Lodge and F.~S. Bates, \enquote{Bridge to loop transition in a
  shear aligned lamellae forming heptablock copolymer,} Macromolecules
  \textbf{37}, 8184 (2004).

\bibitem[Wu \emph{et~al.}(2005)Wu, Lodge and Bates]{re:wu05}
Wu, L., T.~P. Lodge and F.~S. Bates, \enquote{Effect of block number on
  multiblock copolymer lamellae alignment under oscillatory shear,} J. Rheol.
  \textbf{49}, 1231 (2005).

\bibitem[Yih(1967)Yih]{re:yih67}
Yih, C.~S., \enquote{Instability due to viscosity stratification,} J. Fluid
  Mech. \textbf{27}, 337 (1967).

\bibitem[Yih(1968)Yih]{re:yih68}
Yih, C.~S., \enquote{Instability of unsteady flows or configurations: Part 1.
  {Instability} of a horizontal liquid layer on an oscillating plane,} J. Fluid
  Mech. \textbf{31}, 737 (1968).

\end{thebibliography}
\end{document}